\documentclass[12pt,preprint]{aastex}
\begin{document}
\title{Toward better simulations of planetary nebulae luminosity functions 
}
\author{R. H. M\'endez\altaffilmark{1}, A. M. Teodorescu\altaffilmark{1},
   D. Sch\"onberner\altaffilmark{2}, R. Jacob\altaffilmark{2}, 
   M. Steffen\altaffilmark{2}
}
\altaffiltext{1}{Institute for Astronomy,
      University of Hawaii, 2680 Woodlawn Drive, Honolulu, HI 96822}
\email{mendez@ifa.hawaii.edu, ana@ifa.hawaii.edu}
\altaffiltext{2}{Astrophysikalisches Institut Potsdam,
      An der Sternwarte 16, 14482 Potsdam, Germany}

\begin{abstract}

We describe a procedure for the numerical simulation of the planetary 
nebulae luminosity function (PNLF), improving on previous work
(M\'endez \& Soffner 1997). Earlier
PNLF simulations were based on an imitation of the observed distribution 
of the intensities of [O III] $\lambda$5007 relative to H$\beta$, generated 
predominantly using random numbers. We are now able to replace this by a
distribution derived from the predictions of hydrodynamical PN models
(Sch\"onberner et al. 2007),
which are made to evolve as the central star moves across the HR diagram,
using proper initial and boundary conditions. In this way we move
one step closer to a physically consistent procedure for the generation
of a PNLF. As an example of these new simulations, we have been able to
reproduce the observed PNLF in the Small Magellanic Cloud. 

\end{abstract}

\keywords{galaxies: distances and redshifts --- 
galaxies: individual (SMC) --- methods: numerical --- 
planetary nebulae: general --- stars: AGB and post-AGB}

\section{Introduction}

We are quickly approaching the 20th anniversary of the introduction of
the planetary nebulae luminosity function (PNLF) as a tool for extragalactic 
distance determinations (Jacoby 1989; Ciardullo et al. 1989). PNLF
distances are among the most reliable from an empirical point of view,
having been extensively tested (Ciardullo 2003). The only disadvantage
of this method is our own inability to understand how the bright end
of the PNLF can be so bright in stellar populations like those of
elliptical galaxies, where there is no clear evidence of recent star
formation. Theoretical attempts to model the PNLF of an old stellar
population, corresponding to what we expect to find in elliptical galaxies,
assuming single-star post-AGB evolution, have not been successful 
(Marigo et al. 2004; Ciardullo 2006). A possible explanation could involve
massive central stars in old populations produced through binary evolution;
e.g. Ciardullo et al. (2005). Although there is no consensus about the
solution to this problem, its existence underlines the importance of the
PNLF as a probe of late stages of stellar evolution in different stellar
populations.

In view of this potential, it is important to develop a satisfactory
procedure for the numerical simulation of the PNLF. It could seem that,
since we lack a thorough theoretical understanding of the generation of a 
PNLF, modeling it is impossible. However, we expect to show that it
can be done, at least well enough to help in the interpretation 
of the observed PNLFs in many galaxies.

The analytical approximation to the PNLF, as defined by Ciardullo 
et al.\ (1989), although adequate for distance determinations, 
cannot be used for our purposes, because it is defined 
to be fixed and universal; its shape is not affected by
any dependence on stellar population properties. What we want 
is a PNLF based as much as possible on a physically realistic, although 
necessarily simplified, representation of post-AGB evolution.

In the present work we would like to report recent progress in such a 
modeling. Section 2 briefly reviews earlier efforts (M\'endez et al. 1993;
M\'endez and Soffner 1997). Sections 3 and 4 describe new nebular models 
(Sch\"onberner et al. 2007) and how they can be used for PNLF simulations.
The results are shown and discussed in Section 5, and Section 6 gives a
summary and some perspectives for future work.

\section{Early modeling of the PNLF by Monte Carlo methods}

We will present a summary of the procedure used by M\'endez et al. (1993) 
and M\'endez \& Soffner (1997) for the numerical simulation of a PNLF.
Please refer to those papers for more details.

\subsection{Post-AGB ages and masses}

The first step is to generate a set of central stars with random post-AGB 
ages and masses. The post-AGB ages are given by a uniform random distribution 
from 0 to 30000 years. These ages are counted from the moment when the
post-AGB star reaches a surface temperature of 25000 K. The central star
mass distribution starts near 0.55 $M_{\odot}$ because less massive stars 
are expected to evolve too slowly away from the AGB; the nebula is 
dissipated before the remnant star becomes hot enough to ionize it. 
The mass distribution has a maximum around 0.57 $M_{\odot}$ and it 
decreases exponentially for higher masses, to fit the observed white 
dwarf mass distribution in our Galaxy. This random central star mass 
distribution can then be truncated at a certain 
``maximum final mass'' to simulate populations without recent massive star 
formation: all stars more massive than the maximum final mass have already 
evolved into white dwarfs.

Here we have a problem. Where should we truncate? We find empirically 
that, to produce a sufficiently bright PNLF, we must truncate 
at the relatively high mass of 0.63 $M_{\odot}$. But this
mass would seem to be excessively high for old populations like those
in elliptical galaxies. The problem has been described by Ciardullo et al.
(2005). Briefly, if we adopt the initial-final mass relation as empirically
determined by Weidemann (2000), then a final mass of 0.63 $M_{\odot}$
leads inescapably to an initial mass of about 2 $M_{\odot}$. Such stars
do not have pre-white-dwarf lifetimes long enough for us to expect them to 
be still producing PNs in old populations like those of elliptical galaxies. 
Therefore, if we want to explain the bright end of the PNLF in old
populations, we need to explain the origin of the most massive central 
stars. This problem does not have a definitive solution yet, although
mergers of binary systems (blue stragglers, Ciardullo et al. 2005) could
be a possible alternative. Another possible alternative could perhaps be 
a sufficiently wide initial-to-final mass relation, allowing lower
initial masses to sometimes contribute high enough final masses. This 
idea is somewhat unpopular but has not been empirically rejected yet
(Weidemann 2000; Alves et al. 2000; Ferrario et al. 2005). Metallicity 
could certainly play a role in widening the initial-to-final mass 
relation (Meng et al. 2007).

Our position concerning this problem is very simple: we are only 
trying to model the PNLF, not to explain it. We have quite clear evidence 
that massive central stars in old populations exist. The bright end
of the PNLF requires the existence of very luminous central stars, at 
least 7000 $L_{\odot}$. Spectral analyses of the bright PNs confirm this;
see e.g.\ Jacoby and Ciardullo (1999) on the M 31 bulge PNs, and
M\'endez et al.\ (2005) on the PNs in the elliptical galaxy NGC 4697.
Unless there is something terribly wrong with the luminosity--core mass
relation, such luminous central stars have to be more massive than 0.6
$M_{\odot}$.

In addition to those, we can mention a case much closer to us, namely  
the central star of K 648 in the globular cluster M 15. This central star
is bright enough to permit a good non-LTE model atmosphere analysis of its 
absorption-line spectrum, which gives information about its effective
temperature and surface gravity (e.g. McCarthy et al.\ 1997). Together with
the known distance to the cluster, this permits to obtain the luminosity 
and (again using the luminosity--core mass relation) the mass of the
central star, which turns out to be 0.6 $M_{\odot}$ (Alves et al.\ 2000).

In view of the evidence, we adopt the truncation at 0.63 $M_{\odot}$ as
empirically given, and proceed with our numerical simulation.

In the present work we will only refer in passing to the problem of 
the more massive 
central stars expected in populations with abundant recent star formation
(typical examples are the Magellanic Clouds). There are several possible 
mechanisms that can limit the [O III]$\lambda$5007 flux of PNs with
more massive central stars, for example circumstellar extinction. This
has been very well explained in Section 2.1 of Ciardullo et al. (2005), 
so we do not need to repeat it here. The available evidence indicates 
that the truncation near 0.63 $M_{\odot}$ works well enough for all 
population ages.

\subsection{Central star luminosities and surface temperatures}

Having generated random numbers that give post-AGB ages and central star 
masses, we derive for each central star the corresponding luminosity and 
effective temperature, using H-burning post-AGB evolutionary tracks by 
Sch\"onberner (1989) and Bl\"ocker (1995) to build a look-up table and 
an associated bilinear interpolation procedure. For example,
Figure 2 in M\'endez \& Soffner (1997) shows the resulting values of 
luminosity and temperature for the central stars of randomly generated
PNs.

\subsection{Nebular H$\beta$ luminosities and UV photon leaking}

Knowing $L$ and $T_{\rm eff}$, we calculate, using recombination theory,
the H$\beta$ luminosity that the nebula would emit if it were completely
optically thick in the H Lyman continuum. Then we generate a random number, 
subject to several conditions (derived from observations of Galactic PNs 
and their central stars; see next subsection and M\'endez et al. 1992), 
for the absorbing 
factor $\mu$, which gives the fraction of stellar ionizing luminosity 
absorbed by the nebula. We use the absorbing factor to correct the nebular
H$\beta$ luminosity for the effect of UV photon leaking. We consider
the factor $\mu$ to be essential for a successful PNLF simulation, for two 
reasons. First, after the Hubble Space Telescope images, we know that most 
PNs show equatorial density enhancements, suggesting that even if they are 
optically thick in the direction of the equator, they are likely to 
start leaking UV ionizing radiation through the poles very soon. Second, 
we can show (M\'endez \& Soffner 1997) that a PNLF generated under the 
assumption that all PNs are completely optically thick in all directions 
turns out to be too bright. Can we reduce the maximum final mass, instead
of allowing for UV photon leaking? No, because such massive central stars
are known to exist; their suppression is not an option. Can we attribute
the weakening to circumstellar dust extinction? The answer to this question 
is more complicated. Circumstellar dust extinction is probably a dominant 
factor for the most massive central stars in regions with recent star 
formation; as we mentioned at the end of subsection 2.1, this probably 
helps to understand why the bright end of the PNLF is not substantially
brighter than in galaxies without recent star formation. However, we should
expect circumstellar dust extinction to become less important as we consider 
less massive central stars. These less massive central stars are expected to 
evolve more slowly, giving time for the ejected material to dissipate.

At this point we need to introduce the observed behavior of the 
recombination line H$\beta$. Consider the PNs with the brightest
H$\beta$ fluxes in the Magellanic Clouds, as shown in Figure 4a of
Dopita et al. (1992). Some of them are of low excitation class, which
indicates central star surface temperatures around 30,000 K. We know that,
for constant luminosity, the number of H-ionizing photons from the central 
star increases roughly by a factor 2.5 as we go from 
$T_{\rm eff}$ = 30,000 K to 70,000 K. The nebular H$\beta$ luminosity is
nearly proportional to the number of H-ionizing photons. For that reason 
we expect a completely optically thick nebula to show an increasing 
H$\beta$ luminosity as its central star heats up. If we want to keep the 
low-excitation PNs among the brightest in H$\beta$, we need 
increasing UV photon leaking at higher $T_{\rm eff}$. 
Note that here circumstellar dust extinction does not help, 
because we expect more extinction at lower $T_{\rm eff}$ and less 
extinction as the central star heats up and the nebula expands. We conclude
that, in the case of the Magellanic Clouds PNs, it is the absorbing factor, 
not circumstellar dust extinction, that plays a predominant role. We assume
that this conclusion applies in general. Of course the only way to test
this assumption is to obtain deep spectrophotometry of many PNs in 
different galaxies, which we hope can be done in the not too distant 
future. Note that for this purpose the search technique must be oriented 
to detecting PNs in a recombination line like H$\alpha$ or H$\beta$, not 
just those with strong [O III] emission, which of course will never 
belong to low excitation classes.
For a more detailed discussion on the interpretation of H$\beta$
luminosities, please refer to section 6 ``Consistency checks'' in 
M\'endez \& Soffner (1997).

\subsection{More about the absorbing factor}

For easier reference, we repeat here some information given in previous
papers. The empirical basis for the assignment of absorbing factors is a 
study of optical thickness in Galactic PNs (M\'endez et al. 1992). We
generate absorbing factors $\mu$ using random numbers. 
For $T_{\rm eff}$s between
25,000 and 40,000 K, $\mu$ follows a random uniform distribution between 
0.4 and 1.4, with all values higher than 1 replaced by 1. This produces 
a certain predominance of completely optically thick objects, as observed
in Table 4 of M\'endez et al. (1992), but allows for the observed fraction
of optically thin PNs with low-$T_{\rm eff}$ central stars. 
For $T_{\rm eff}$s between 40,000 K and the beginning of the white dwarf 
cooling track, $\mu$ follows a random uniform distribution between 0.05
and the parameter $\mu_{\rm max}$. We adopt $\mu_{\rm max} = 1$; in this 
way some percentage of the bright PNs with very hot central stars can have
$\mu$ close to 1.
For central stars on the white dwarf cooling tracks, $\mu$ is set equal
to a random number uniformly distributed between 0.1 and 1, and this number 
is multiplied by a factor (1$-$(age(years)/30,000)). In this way we ensure
that $\mu$ tends to 0 as the nebula dissipates.

We have kept the random generation of $\mu$ as simple as possible,
because the amount of empirical information is quite limited.
There is no explicit influence of the central star mass, for example,
basically because we lack credible empirical information that could
guide our modeling. It will always be possible to complicate the 
computer codes once more information becomes available. For the moment 
our simple procedure appears to work well. Although our physical 
interpretation of the absorbing factor is open to future refinements,
we would like to emphasize that once we introduce the
absorbing factor, as constrained by the information we have about
optical thickness of PNs in our Galaxy, the PNLF we generate agrees 
with the observed ones, without any further adjustment.

\subsection{The intensity ratios [O III]$\lambda$5007/H$\beta$}

Since we have generated the H$\beta$ luminosities, now we only need to 
generate the ratios $\lambda$5007/H$\beta$ to obtain the $\lambda$5007
luminosities and compute the PNLF. At this point we depart from M\'endez
\& Soffner (1997). They used mostly random numbers to generate the
intensity ratios, in such a way that the observed histograms of 
$\lambda$5007/H$\beta$ ratios in our Galaxy and the LMC could be 
approximately reproduced. Instead, we want to calculate our 
$\lambda$5007/H$\beta$ ratios from hydrodynamical PN models
(Sch\"onberner et al. 2007). Several evolutionary sequences
of model PNs have been constructed, one sequence for each of a
limited number of central star masses. In the following sections 
we briefly review the basic characteristics of these models, and 
we explain the interpolation procedure we have implemented to obtain 
$\lambda$5007/H$\beta$ ratios for any combination of post-AGB age and
central star mass. In this way we move one step closer to a physically 
consistent procedure for the generation of a PNLF. 

\section{Modeling the PN evolution}

The PN model sequences produced by Sch\"onberner et al. (2007,
in what follows SJSS07)
are based on coupling a spherical circumstellar envelope, assumed 
to be the relic of a strong AGB wind, to a H-burning post-AGB model, 
and following the evolution of the whole system across the H-R 
diagram toward the white-dwarf cooling track. The goal is to produce
radiation-hydrodynamics simulations with the proper initial and
boundary conditions. A one-dimensional 
radiation-hydrodynamics code is employed (Perinotto et al. 1998).
This code is designed to compute ionization, recombination, heating,
and cooling, fully time-dependently. The chemical composition is 
typical for Galactic disk PNs (slightly below solar; see Table 1 in
SJSS07). Nebular evolutionary sequences have been 
computed for central stars with masses 0.565, 0.585 (unpublished), 
0.595, 0.605, 0.625 and 0.696 $M_{\odot}$. Their corresponding 
post-AGB evolutionary tracks in the HR diagram are shown in Figure 1.

Although the models are spherically symmetric, they represent the 
observed nebular structures, as indicated by the H$\alpha$ brightness 
distributions, extremely well. These nebular models show in many cases
a transition between optically thick and thin in the Lyman continuum.
Note, however, that it is not clear if the models can accurately predict 
what fraction of the H-ionizing radiation is being lost through the
nebular poles, due to departures from spherical symmetry in the real 
nebulae. At this point it looks better to use the absorbing factors 
$\mu$ as defined by M\'endez et al.\ (1992, 1993), and combine them
with {\it intensity ratios} $\lambda$5007/H$\beta$, which are not 
too much affected by the onset of UV photon leaking; see Figure 2. 
We believe that 
a combination of spherically symmetric nebula plus $\mu$ absorbing
factor may be a good compromise to describe the evolution of PNs
in a more realistic way than previously attempted, without having to
introduce the enormous complexities of two-dimensional hydrodynamics.

In summary, here we use the SJSS07 model sequences
for one purpose only: to obtain the intensity ratio 
$\lambda$5007/H$\beta$ as a function of nebular post-AGB age.
The resulting run of this ratio for the six central star masses is 
shown in Figure 3. Next step is to implement an interpolation procedure 
that will provide similar information for any central star mass.

\section{Interpolation method for the generation of $\lambda$5007/H$\beta$}

To begin with, we have a table giving central star surface temperature 
$T$, central star luminosity $L$, and nebular ratio $\lambda$5007/H$\beta$, 
which we will call $R$, as a function of post-AGB age $t$, for each of the
six central star masses listed above. The interpolation between these 
tracks is done following a technique described by van der Sluys et al.
(2005). We first divide each of the nebular evolutionary tracks shown in 
Figure 3 into three sections: (a) where $R$ increases until it 
reaches a maximum; (b) where $R$ decreases; (c) where $R$ stabilizes.
There is one exception: since the transition between 0.595 and 0.585
$M_{\odot}$ is somewhat different, in that case we have modified (b) and 
(c) in the following way: (b) where $R$ is between the two peaks; 
(c) where $R$ decreases.

For each of these 3 sections we define a path length $l$ by the following 
expression:

\begin{equation}
l=\sum_{i} \sqrt{\Big(\frac{t(i)-t(i-1)}{\Delta t} \Big)^{2} 
+ \Big(\frac{R(i)-R(i-1)}{\Delta R} \Big)^{2} }
\end{equation}

In this equation, $i$ is the index corresponding to the successive data
rows in each table, and the quantities $\Delta t$ and $\Delta R$ are the
total increments in $t$ and $R$ between the beginning and end of each 
section. 

Each of these 3 sections is redistributed into a fixed number of data
points, equally spaced in the path length. The values for these equally 
spaced points are calculated by polynomial interpolation along each track.
Having done this, each section of each nebular evolutionary track has the
same number of data points, and one point in any section, like (a) to
fix ideas, marks an evolutionary state similar to that of the same point 
along the (a) section in any other nebular evolutionary track.

Now we are able to interpolate between adjacent nebular evolutionary 
tracks, building point by point a new track for each randomly generated 
central star mass. Figure 3 shows two simulated tracks, in the $R$-$t$
plane, produced with this interpolation technique. Their corresponding 
central star
post-AGB evolution in the HR diagram is also plotted in Figure 1. Once
in posession of the time evolution of $R$ for any randomly generated 
central star mass, we can obtain $R$ for the randomly generated 
post-AGB age, and we can proceed to build the PNLF.

\section{Results and discussion}

Our ultimate purpose is to generate a physically consistent PNLF, 
eliminating as much as possible the random numbers used in previous 
modeling, which were reflecting our lack of information about the 
evolution and properties of the PNs at each specific moment. 
At the present time we cannot produce a fully satisfactory simulation, 
because we would need first to explore variations in many input 
parameters and their effect on the PN evolution. For example, we cannot
discuss metallicity effects until we have nebular evolutionary tracks 
for a broad range of metallicities. But we would like to show the very
promising results of a simulation based on the limited number of 
nebular evolutionary tracks presented by SJSS07.

First of all we consider the histogram of the intensity of 
[O III]$\lambda$5007, on the scale $I({\rm H}\beta)=100$. On 
this scale, $I$(5007) is equivalent to 100 $R$. In previous work, we 
simulated the observed histograms of $I$(5007) in our Galaxy and the
LMC using predominantly random numbers. Can we obtain a satisfactory
fit to the observations using instead the ratios $R$ generated by our
PN evolution programs? Before attempting that, we need to consider 
selection effects: which of our generated PNs would actually be 
observable? We seek guidance in Figure 4, which is a modified version 
of Figure 13 in SJSS07 (we could have used their Figure 14 instead), 
showing the excitation class (defined in Eq. (2) of SJSS07) 
as a function of the nebular absolute magnitude 
$M$(5007). Note that the observed PNs are enclosed by the 
nebular evolutionary tracks corresponding to central star masses of 
0.696 and 0.585 $M_{\odot}$. Nebulae that belong to less massive 
central stars fail to become bright, because the central stars evolve
very slowly away from the AGB, and the nebulae dissipate, becoming very 
optically thin, and displaying a very low surface brightness. This 
indicates that they are probably missing in the observed samples.

Thus, in building the theoretical distribution of $I$(5007),  
in order to be consistent with the nebular properties that 
result from the initial and boundary conditions assumed in
SJSS07, we have decided to eliminate the contribution 
from all central stars less massive than 0.585 $M_{\odot}$.
The central star mass distribution we adopted will be shown 
later (see the upper mass distribution in Figure 8).
We have also eliminated all PNs with central stars fainter
than log $L/L_{\odot}$ = 2.4. This was done, in the same way as in 
M\'endez \& Soffner (1997), in order to compensate for an
obvious selection effect: the observed distributions in our Galaxy
and in the LMC are not likely to include PNs with very low-L central
stars, all of which have high surface temperatures. 

Figure 5 shows, then, our corrected theoretical distribution, compared 
with two observed distributions: one for 118 PNs in the LMC (data taken 
from Wood et al. 1987; Meatheringham et al. 1998; Jacoby et al. 1990;
Meatheringham \& Dopita 1991a, 1991b; Vassiliadis et al. 1992) and
another one for 983 PNs in our Galaxy, taken from the Strasbourg-ESO 
Catalogue of Galactic PNs (Acker et al. 1992). These are the same two 
distributions used in Figure 3 of M\'endez \& Soffner (1997). Our
new distribution provides a quite satisfactory fit. We do not
expect a perfect fit, of course, because there are even differences 
between the two observed distributions, the reasons for which are not 
clear at the present time.
Prompted by the anonymous referee, we also show in Figure 6 that
the nebular model sequences in SJSS07 can predict the observed 
distribution of PNs in a diagram of the [O III] $\lambda$5007 to 
H$\alpha$ + [N II] line ratio as a function of $M$(5007), like the
one shown in Fig. 2 of Ciardullo et al. (2002).

Since we have been able to produce a value of $I$(5007) for every
pair of values of post-AGB age and central star mass in our
simulations, we can proceed to build the new $\lambda$5007
PNLF. Figure 7 shows a
comparison between the old PNLF (M\'endez \& Soffner 1997) and the 
new one. The agreement between the two simulations at the bright 
end is excellent. There is a difference at fainter magnitudes, which
does not affect the use of the PNLF for distance determinations.

What is the nature of the ``camel shape'' apparent in the new 
simulation? It can be described as a relative lack of PNs 
for $M$(5007) between $-3$ and 0. In fact, it was already present 
in the M\'endez \& Soffner (1997) simulations, but it is more 
pronounced here. We believe that the most natural explanation of 
this deficit of PNs at intermediate luminosities is related to the 
fact that the central stars in our simulation are shell H-burners.
See Section 9 in M\'endez (1999). 
Post-AGB evolutionary tracks show a quick drop in luminosity 
as the H-burning shell is extinguished and the star goes into the
white dwarf cooling track. For that reason there is a lack of 
central stars at log $L/L_{\odot}$ below 3.5. This lack of central 
stars at intermediate luminosities can explain the lack 
of intermediate-brightness PNs in the PNLF. 

If this explanation is correct, then it should also explain the
different shapes in Figure 7. The most important difference between
the new and the old simulation is that the old one uses a central
star mass distribution extending down to masses as low as 0.55 
$M_{\odot}$. The luminosity drop suffered by H-burning central 
stars is in fact much less dramatic for lower-mass central stars,
and therefore we expect such low-mass central stars to help reduce
the deficit, as observed in Figure 7.

Let us show this effect in more detail. In Figure 8 we show three
simulations. The first one uses a central star mass distribution 
with a sharp low-mass cut 
at 0.585 $M_{\odot}$. In the 2nd and 3rd cases we allow the mass
distribution to be extended toward less massive central stars. Since
SJSS07 cannot be used at these low masses, because the nebulae are 
predicted to be too faint, for these low-mass central stars we 
used the procedure of M\'endez and Soffner (1997) to generate the 
values of the ratios $\lambda$5007/H$\beta$. Indeed it appears that
the addition of more and more lower-mass central stars increasingly 
reduces the deficit, as expected. Of course we would need to 
investigate if it is possible to impose reasonable initial 
conditions that will result in visible PNs around the lower-mass
central stars. We assume that this is possible, but such an 
investigation is extremely time-consuming and lies outside the 
scope of the present work.

We should mention in passing that another way of decreasing the
deficit is to allow for a certain percentage of He-burning central 
stars, which do not show a quick luminosity drop. We cannot include 
such evolutionary tracks in our procedure; see M\'endez \& Soffner 
(1997) and the discussion (section 7) in SJSS07.

The new PNLF shape we have obtained reminds us immediately of the
observed PNLF in the SMC as described by Jacoby \& De Marco (2002).
Therefore in Figure 9 we fit the Jacoby \& De Marco data with our new 
PNLF. The agreement is very encouraging. The fit to the bright end 
gives a distance modulus of 19.3 mag, which agrees, within the rather 
large uncertainties, with the 19.1 obtained by Jacoby et al. (1990, 
see in particular their Figure 5). Whenever we fit the PNLF we are 
making a simultaneous fit to both 
the distance modulus and to the total number of PNs in the galaxy 
in question; thus our new simulation also implies a total number
of approximately 120 PNs in the surveyed area of the SMC, 
in rough agreement with estimates by Jacoby and De Marco (2002).
Note that the PN numbers observed at faint magnitudes are probably 
affected by some incompleteness; we are fitting only the bright 
end of the SMC PNLF, which appears to be complete, as discussed 
by Jacoby and De Marco. 

In Figure 9 we find that the mass distribution with a sharp cut at 
0.585 $M_{\odot}$ gives a better fit than other distributions that
include a contribution from lower central star masses. A lack of
low-mass central stars in the SMC may have different possible 
interpretations. It might reflect lack of star formation at earlier 
times, producing a lack of the corresponding low initial masses
(Ciardullo et al. 2004);  
it might also mean that, in the SMC, low-mass central stars find it 
more difficult to produce visible PNs, perhaps as a consequence of 
the low metallicity. These ideas will have to be tested 
when nebular evolution models like those of SJSS07, but for lower 
metallicities, become available. The PNLF shape may provide useful 
diagnostics for studies of star formation history and post-AGB 
evolution in different populations.

\section{Recapitulation and perspectives}

We have shown that using models like those of SJSS07 it is 
possible to generate a numerical simulation of a PNLF, if we are 
willing to assume an empirically given central star mass distribution,
which however still needs to be justified from stellar evolution and
population evolution theories. Leaving that problem aside, the new 
procedure is able to reproduce observed histograms of $I$(5007), and
the new generated PNLF agrees with the old one at the bright end, 
which means that it gives the same PNLF distances as before. In 
addition, we have found that the shape of this new simulated 
PNLF explains the observed PNLF shape in the SMC quite well.

What remains to be done is to systematically explore the initial 
parameters controlling the PN model evolution, to see what effects 
they have on the PNLF. In particular, initial conditions may 
influence the low-mass cut we need to apply in the central star 
mass distribution, probably in part as a function of metallicity.

The results we have presented offer some promise for future
PNLF research. Most important, if it is possible to produce new 
PN evolution models for a variety of metallicities (a difficult 
task, because it requires at the very least to have a good 
theoretical treatment of AGB and post-AGB mass loss, in order 
to deal with both central star and nebular evolution), then 
it will become comparatively easy, using the methods described 
here, to investigate metallicity effects on the PNLF. If it is 
possible to build observed PNLFs for several galaxies down 
to fainter magnitudes, 
then the different PNLF shapes, if confirmed, would provide a 
very useful diagnostic for population characteristics like star
formation history, central star mass distribution of observable 
PNs, or perhaps even the relative frequency of He-burners among 
PN central stars.

\acknowledgements
This work has been supported by the National Science Foundation, 
under grant 0307489. We thank the anonymous referee for some useful
comments.

\clearpage

\begin{figure}
\epsscale{1.0}
\plotone{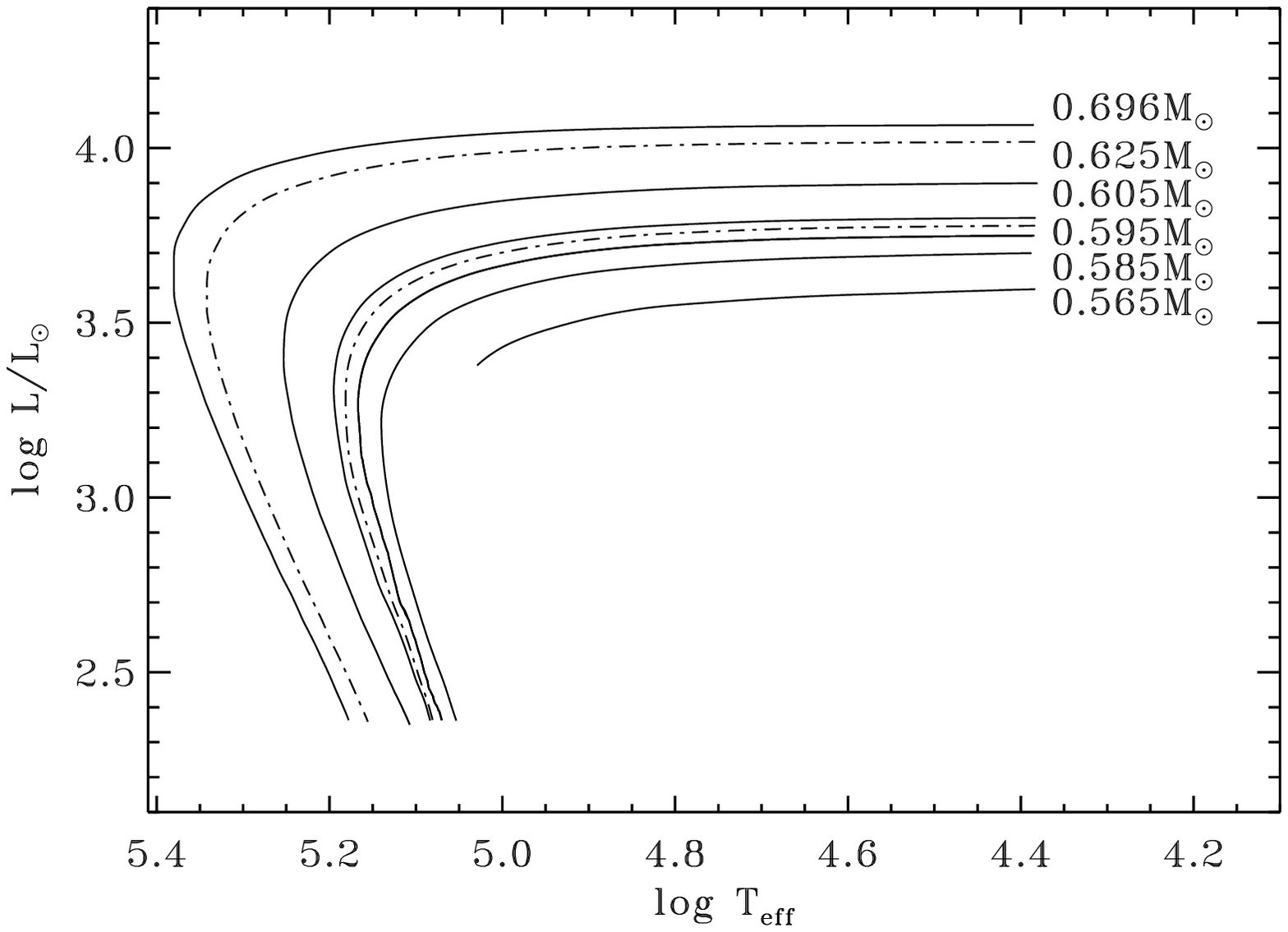}
\caption{
Solid lines are post-AGB evolutionary tracks for six central star masses 
in the log $T_{eff}$--log $L$ plane. Dashed lines (unlabeled) are two 
interpolated tracks generated as in M\'endez \& Soffner (1997).
}
\end{figure}

\begin{figure}
\epsscale{1.0}
\plotone{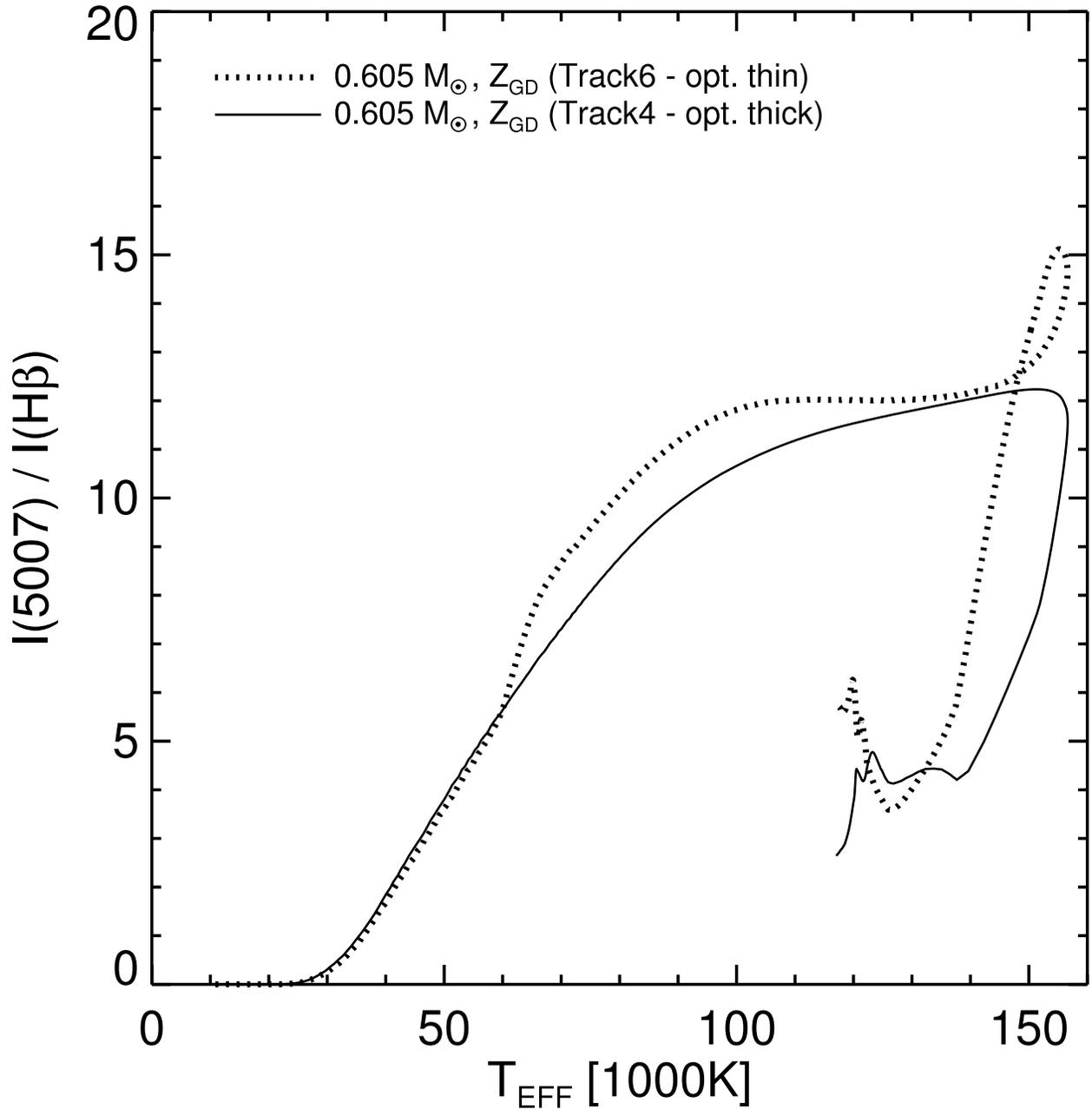}
\caption{
I(5007)/I(H$\beta$) line ratio vs central star effective temperature for 
two hydrodynamical sequences calculated in SJSS07. $Z_{\rm GD}$ means 
the metallicity of our Galactic disk. The nebula following 
Track 4 (solid line) remains always optically thick, while along Track 6 
(dotted line) the nebula becomes optically thin to H-ionizing photons 
as the central star evolves. The ratio is always larger in the optically 
thin phase, but the difference is seldom larger than about 10\% in the 
relevant bright phases.
}
\end{figure}

\begin{figure}
\epsscale{1.0}
\plotone{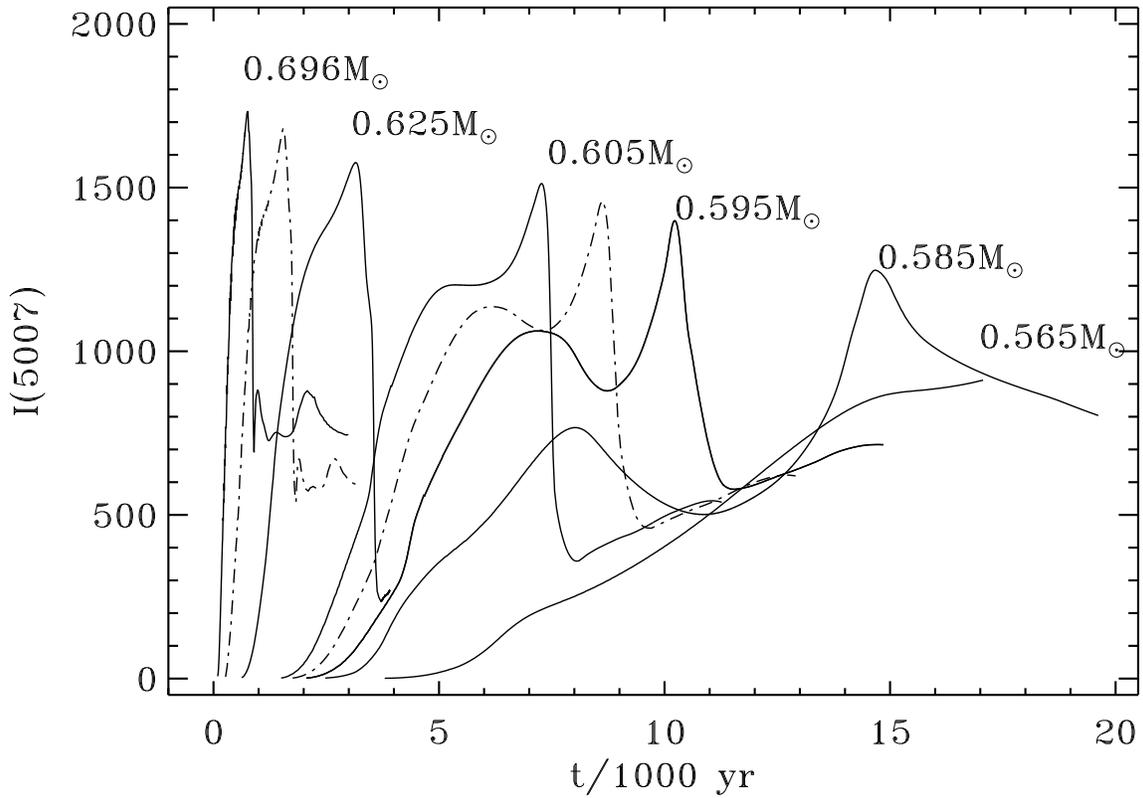}
\caption{
Solid lines are PN evolutionary tracks for the six 
central star masses, taken from SJSS07, in the 
$t$--$I$(5007) plane; $I$(5007) is on the scale $I(H\beta)$=100. 
The dashed lines (unlabeled) are two interpolated tracks, each
corresponding to one of the interpolated stellar evolutionary tracks
shown in Fig. 1. The interpolated PN evolutionary track generation 
is explained in Section 4.
}
\end{figure}

\begin{figure}
\epsscale{1.0}
\includegraphics[bb= -7.cm 10cm 20cm 17cm, width=1.6\textwidth]{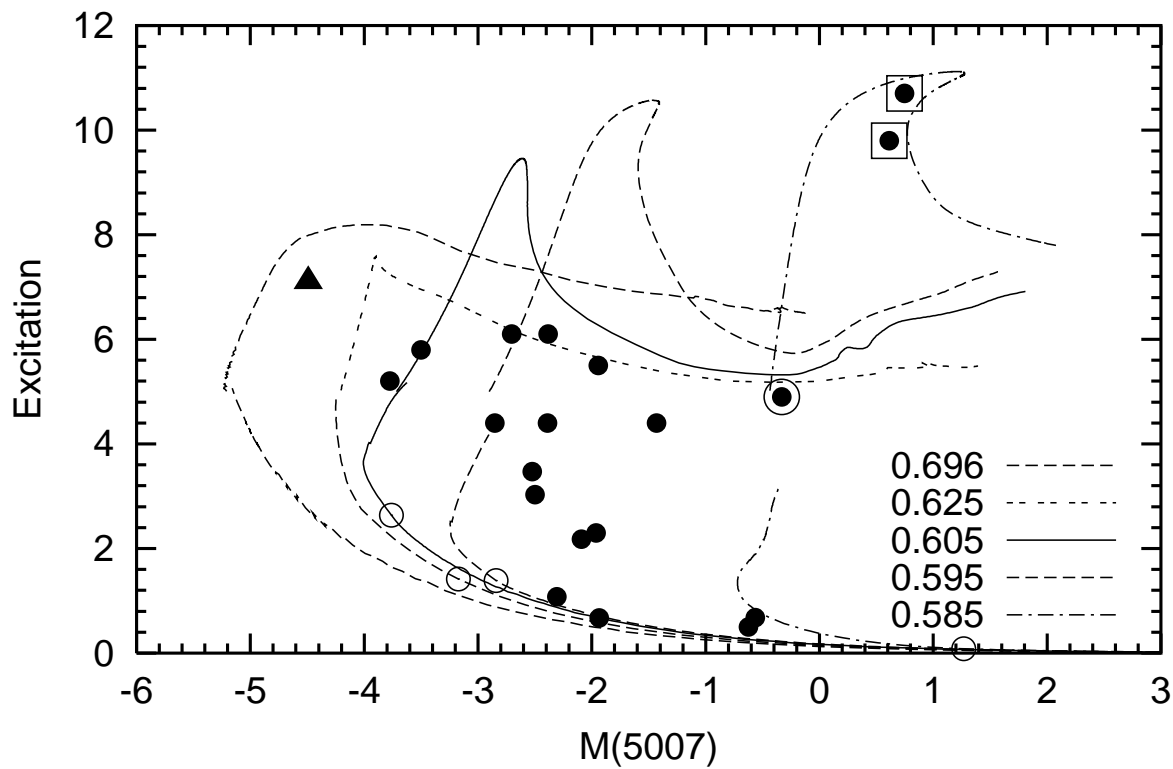}
\caption{
Nebular excitation parameter vs. absolute $\lambda$5007 magnitude, 
for five hydrodynamical sequences labeled according to their mass.
Open circles along the tracks indicate the moment when
the nebular models become optically thin for Lyman continuum photons.
The gaps seen for some tracks are artifacts caused by the definition of
the excitation parameter (see also SJSS07, Figs.\ 13 and 14 therein).
Data of Galactic PNs
with spectroscopically determined distances are shown as dots for 
comparison. This figure is a modified version of Fig. 13 in SJSS07.
The 'circled' dot belongs to
NGC 7293, the two 'squared' dots to the high-excitation PNs 
NGC 1360 and NGC 4361. The filled triangle marks the position
of NGC 7027.
}
\end{figure}

\begin{figure}
\epsscale{1.0}
\plotone{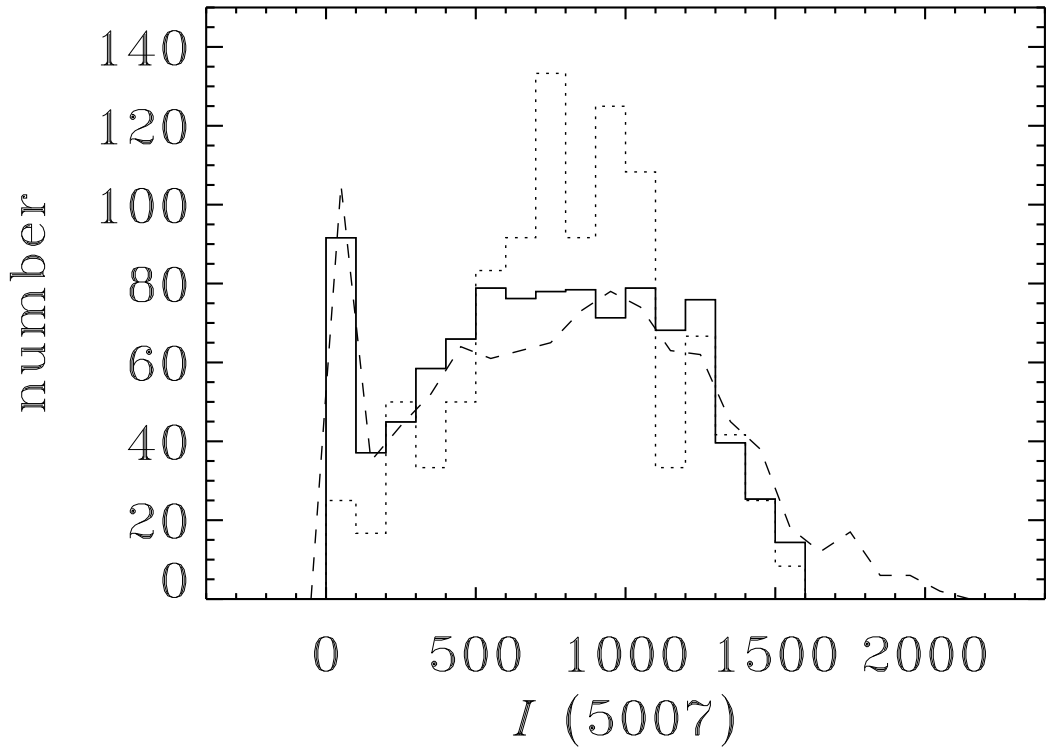}
\caption{
Histograms of the intensity of $\lambda$5007,
on the scale $I(H\beta)$=100. The dashed line indicates the histogram 
for 983 objects in our Galaxy. The other two histograms have been
normalized to this number. The dotted line is the histogram for 118 
LMC objects. The full line is our new distribution, generated
as described in the text.
}
\end{figure}

\begin{figure}
\epsscale{1.0}
\plotone{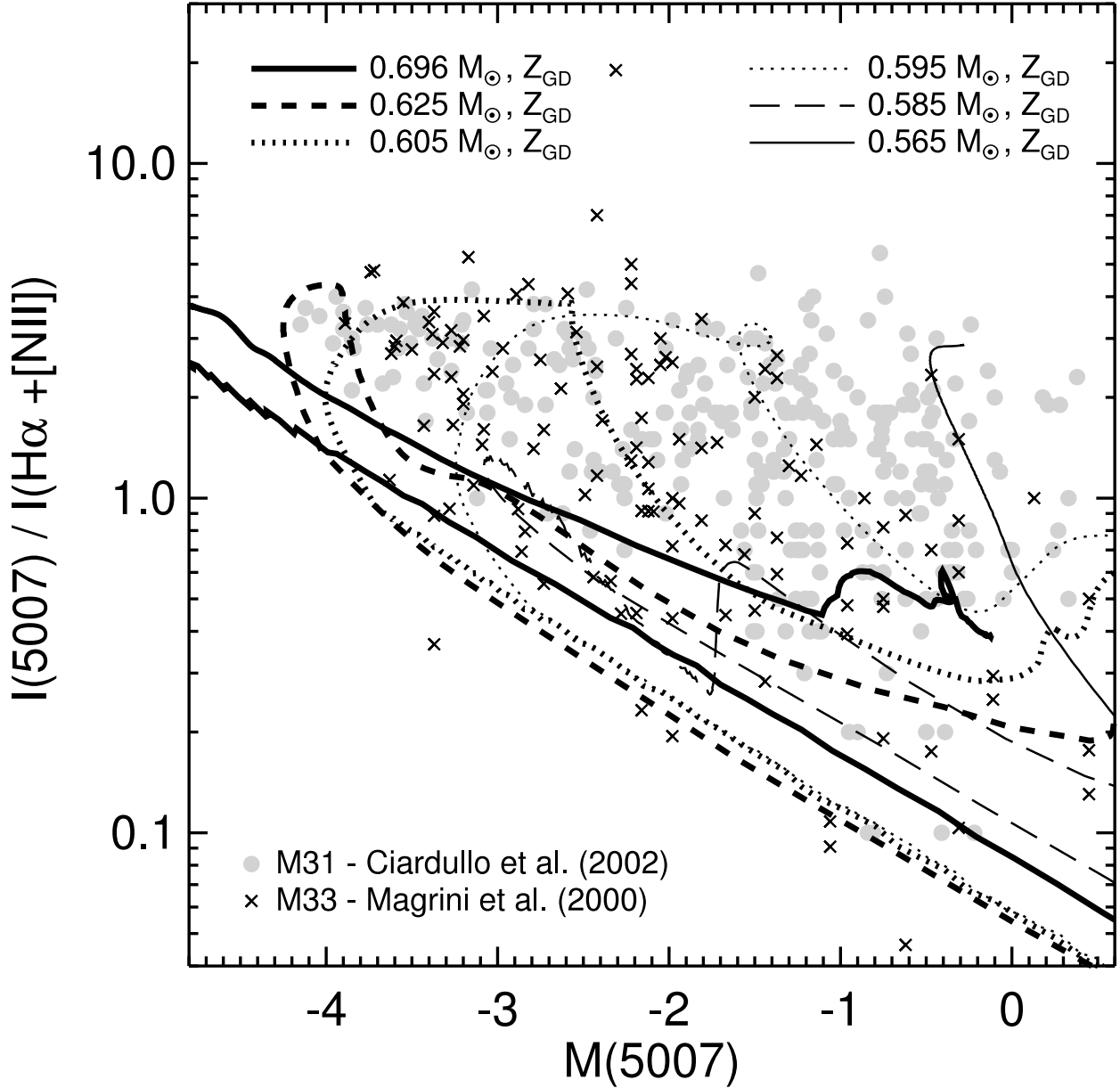}
\caption{
[O III] $\lambda$5007 to H$\alpha$ + [N II] line ratios for 
PNs in the M 31 bulge and M 33. The data, taken from Ciardullo et al. 
(2002) and Magrini et al. (2000), are compared with PN evolutionary
tracks from SJSS07. $Z_{\rm GD}$ means the metallicity of our 
Galactic disk. Evolution is from lower right to upper left and 
back. The models cover the observed range very well; no internal
reddening corrections are needed. The reason for the small number 
of PNs near the upward-moving tracks is that the probability of 
finding them there is low. Please refer to SJSS07 (section 5 and
figure 15). The PN brightens quickly, and then it fades more slowly, 
so that we find most of the observed PNs in the fading region.
}
\end{figure}

\begin{figure}
\epsscale{1.0}
\plotone{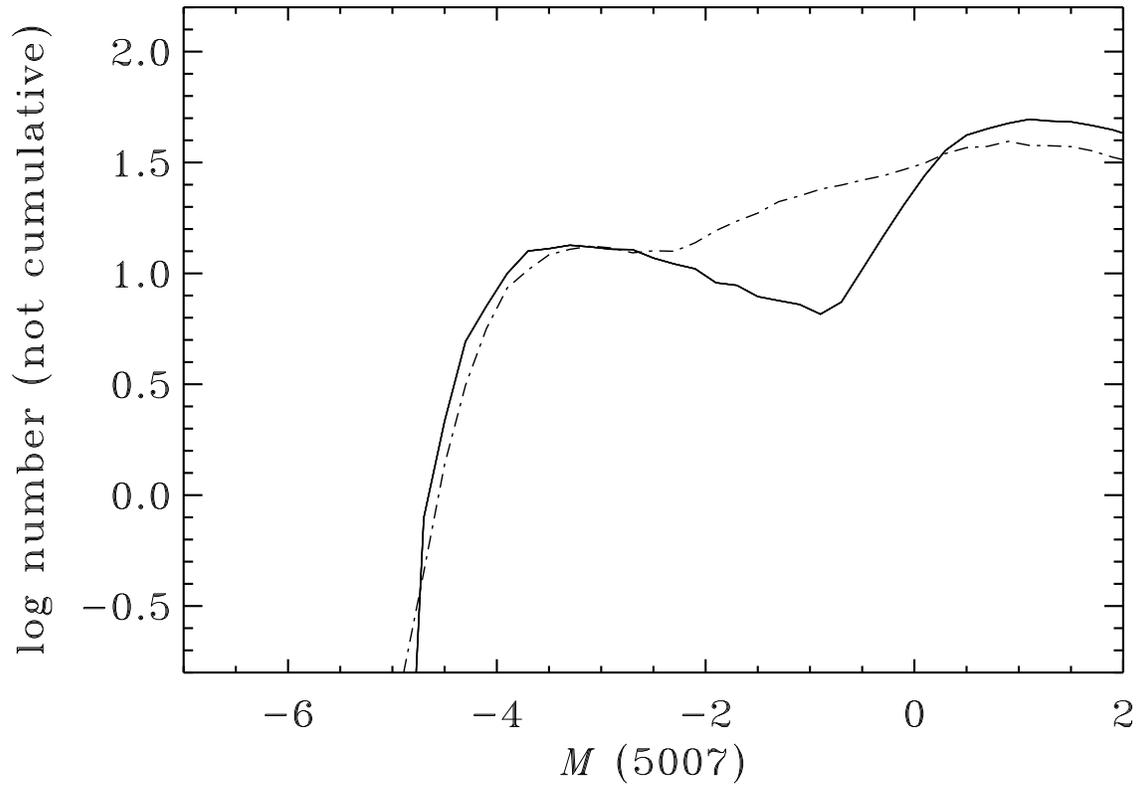}
\caption{
PNLF (full line) generated using the hydrodynamical models
by SJSS07, compared with the old PNLF (dashed
line) generated as described in  M\'endez \& Soffner (1997).
}
\end{figure}

\clearpage
\thispagestyle{empty}
\setlength{\voffset}{-12mm}
\begin{figure}
\epsscale{1.0}
\plottwo{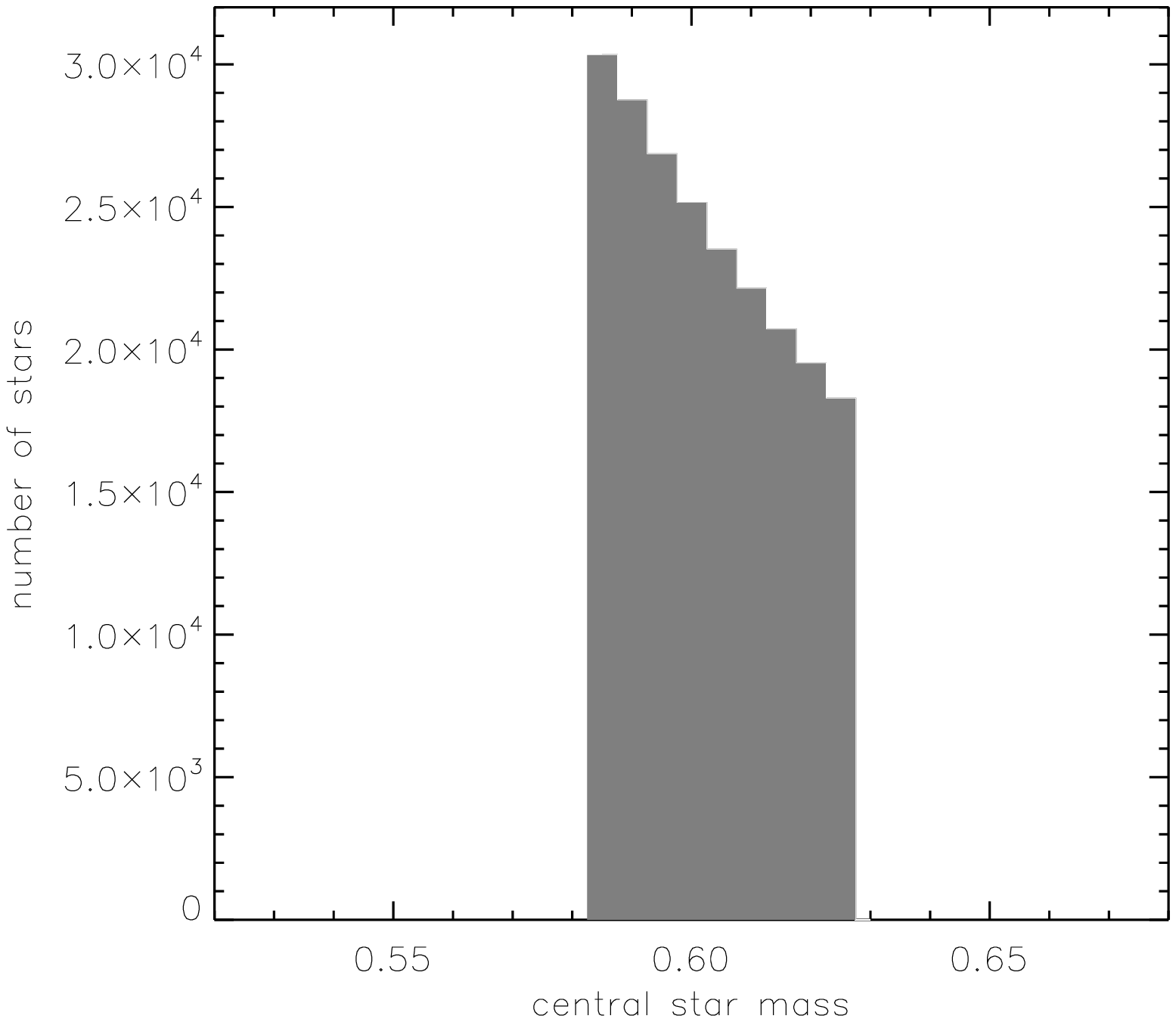}{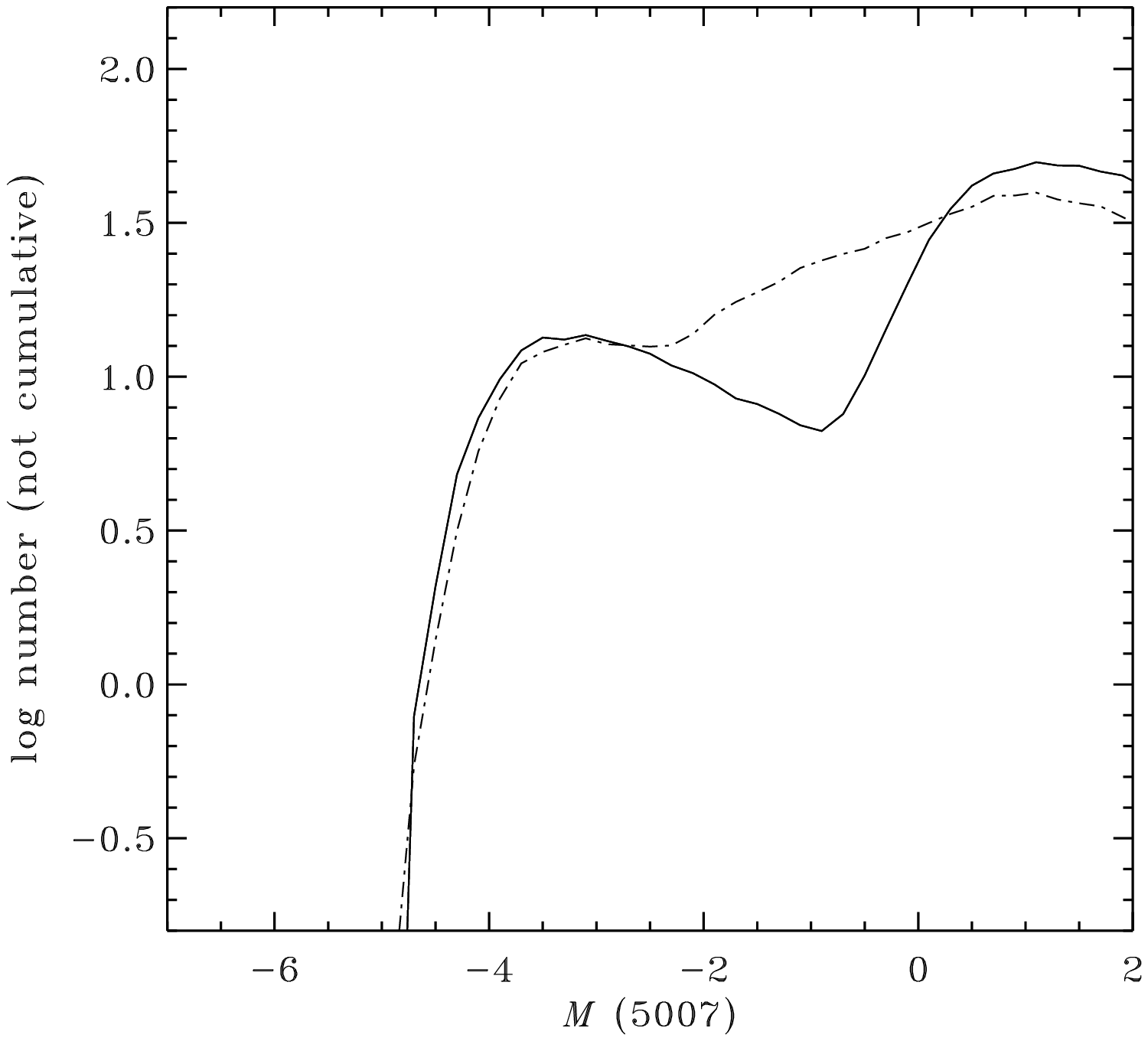}
\plottwo{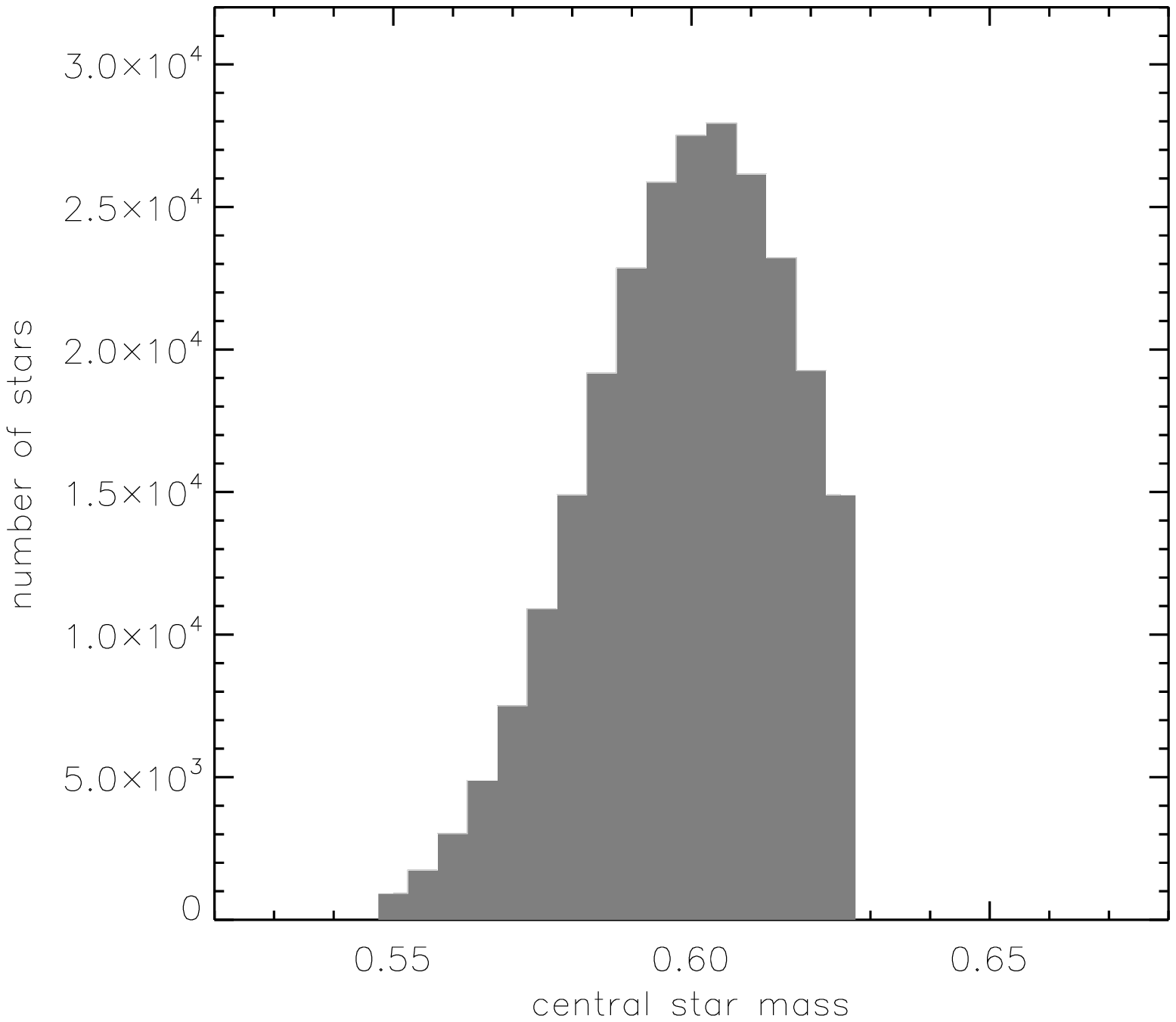}{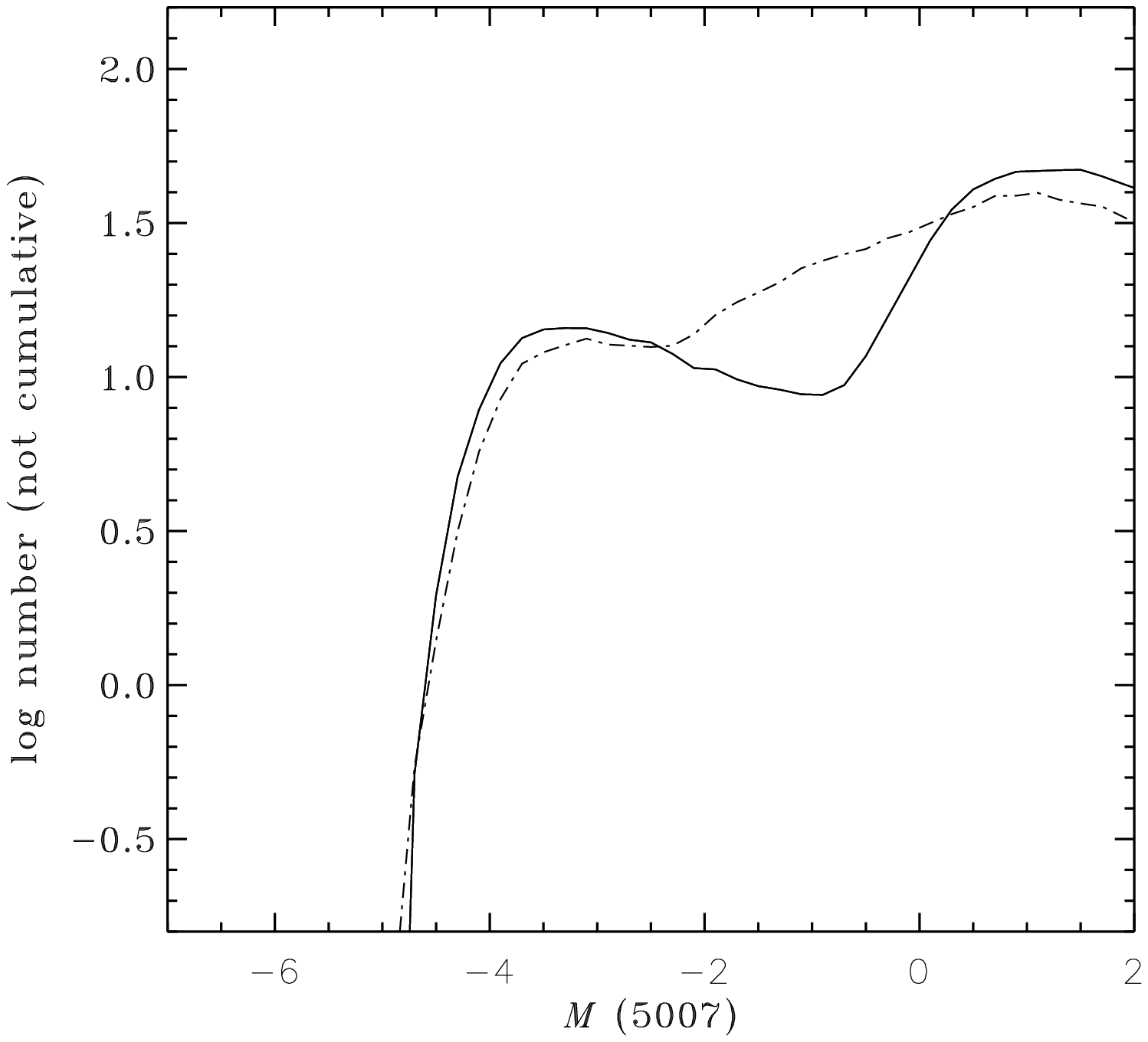}
\plottwo{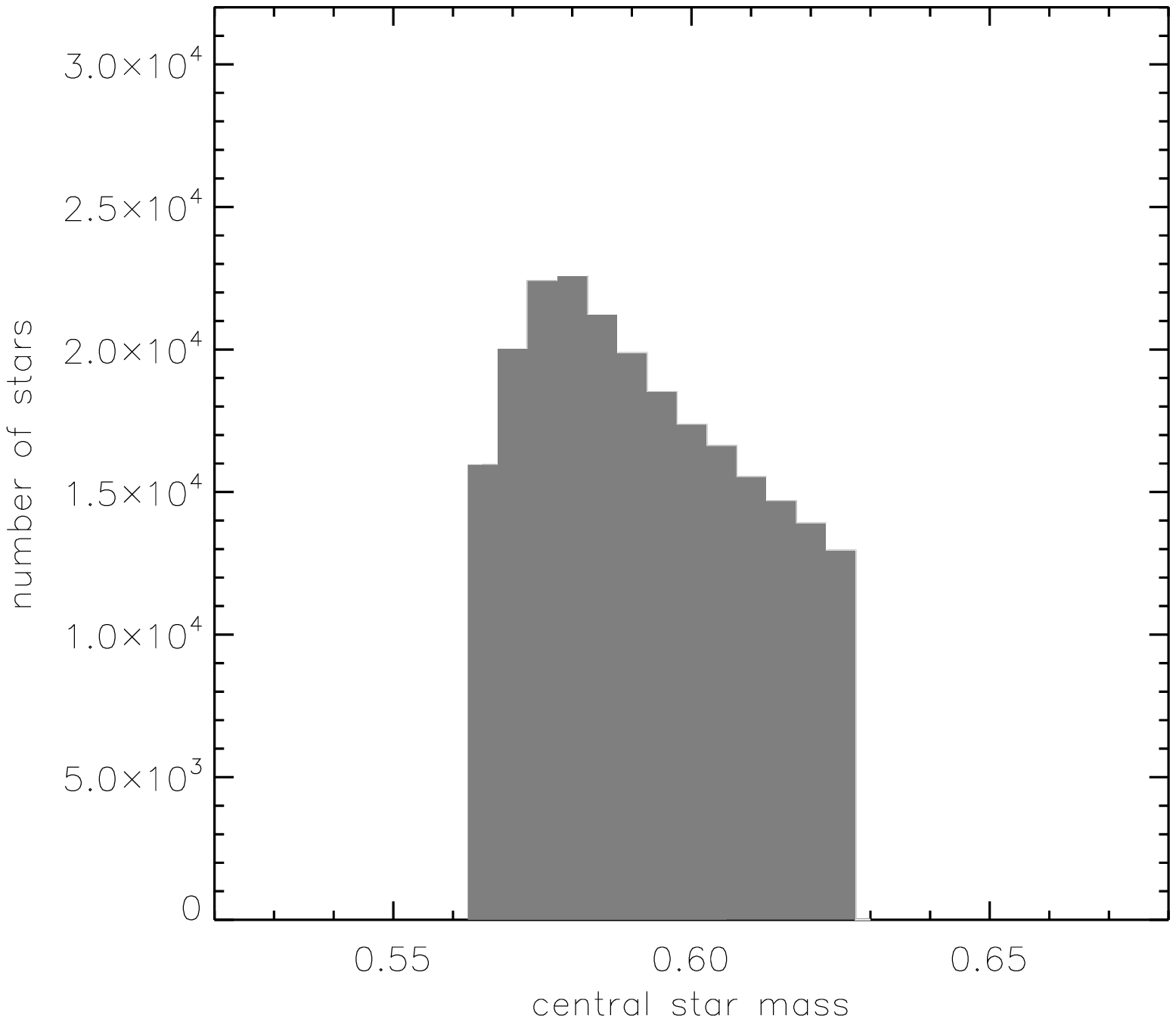}{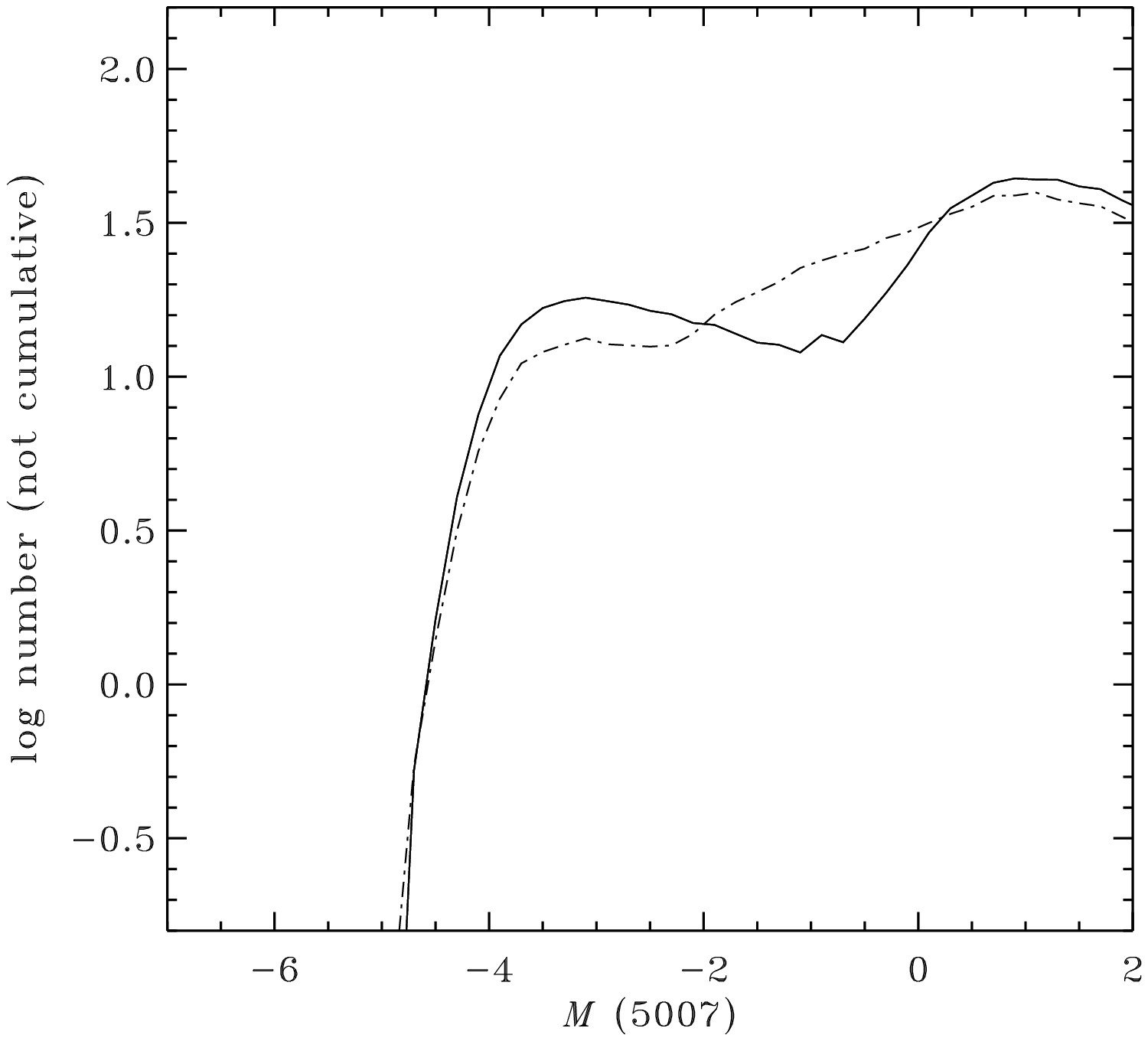}
\caption{
PNLFs generated using different central star mass distributions. 
The mass distributions are on the left, and the corresponding PNLFs 
(full lines) are on the right. The first (upper) mass distribution 
is the one used to produce Fig. 7, with a sharp low-mass cut at 0.585 
$M_{\odot}$. The 2nd and 3rd include progressively 
more low-mass central stars. The three PNLFs are compared with
the old PNLF generated as in M\'endez and Soffner (1997), indicated
with a dash-dotted line. As we increase the number of low-mass stars,
the PNLF becomes more similar to the old one.
}
\end{figure}
\clearpage
\setlength{\voffset}{0mm}

\begin{figure}
\epsscale{1.0}
\plottwo{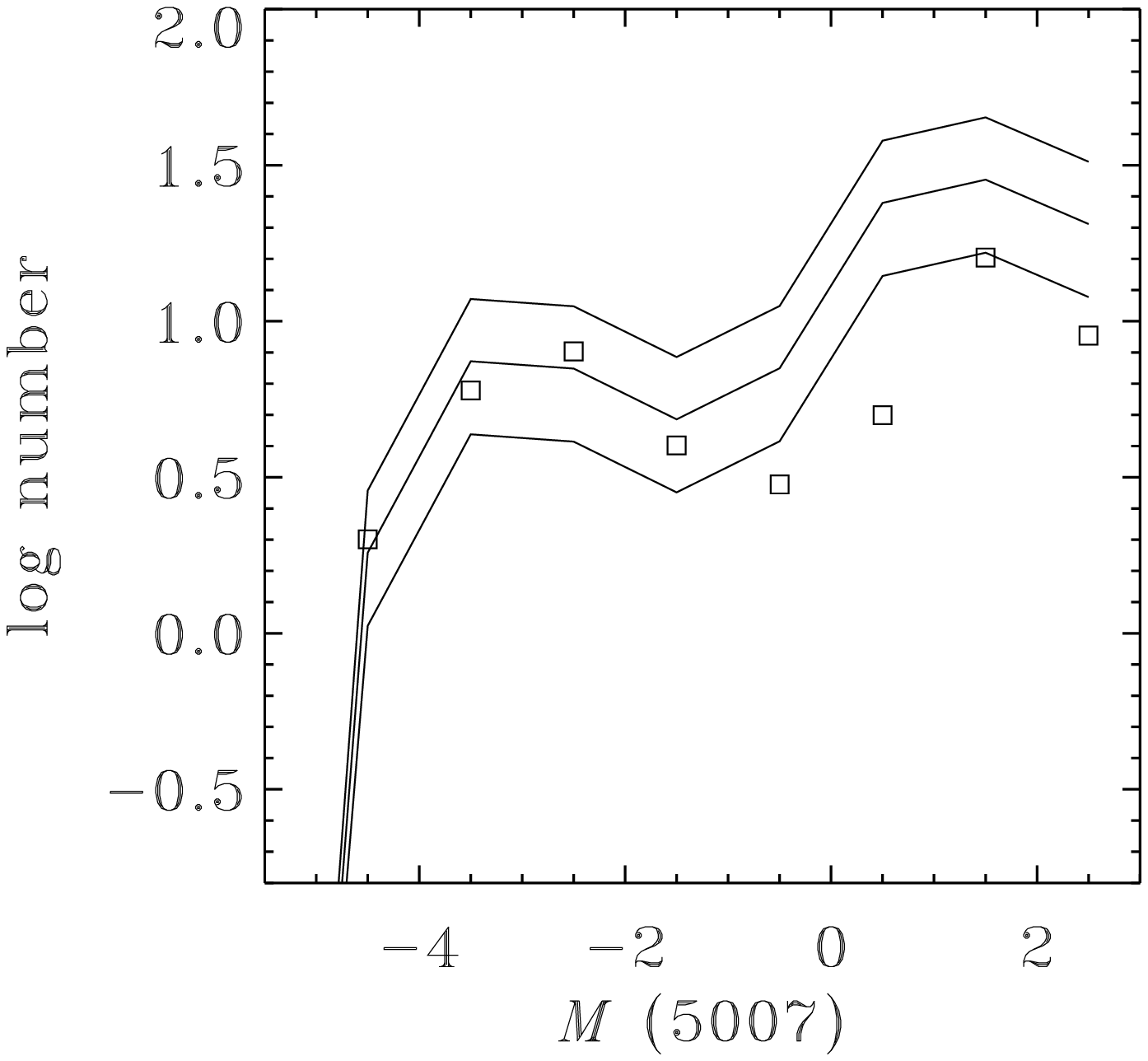}{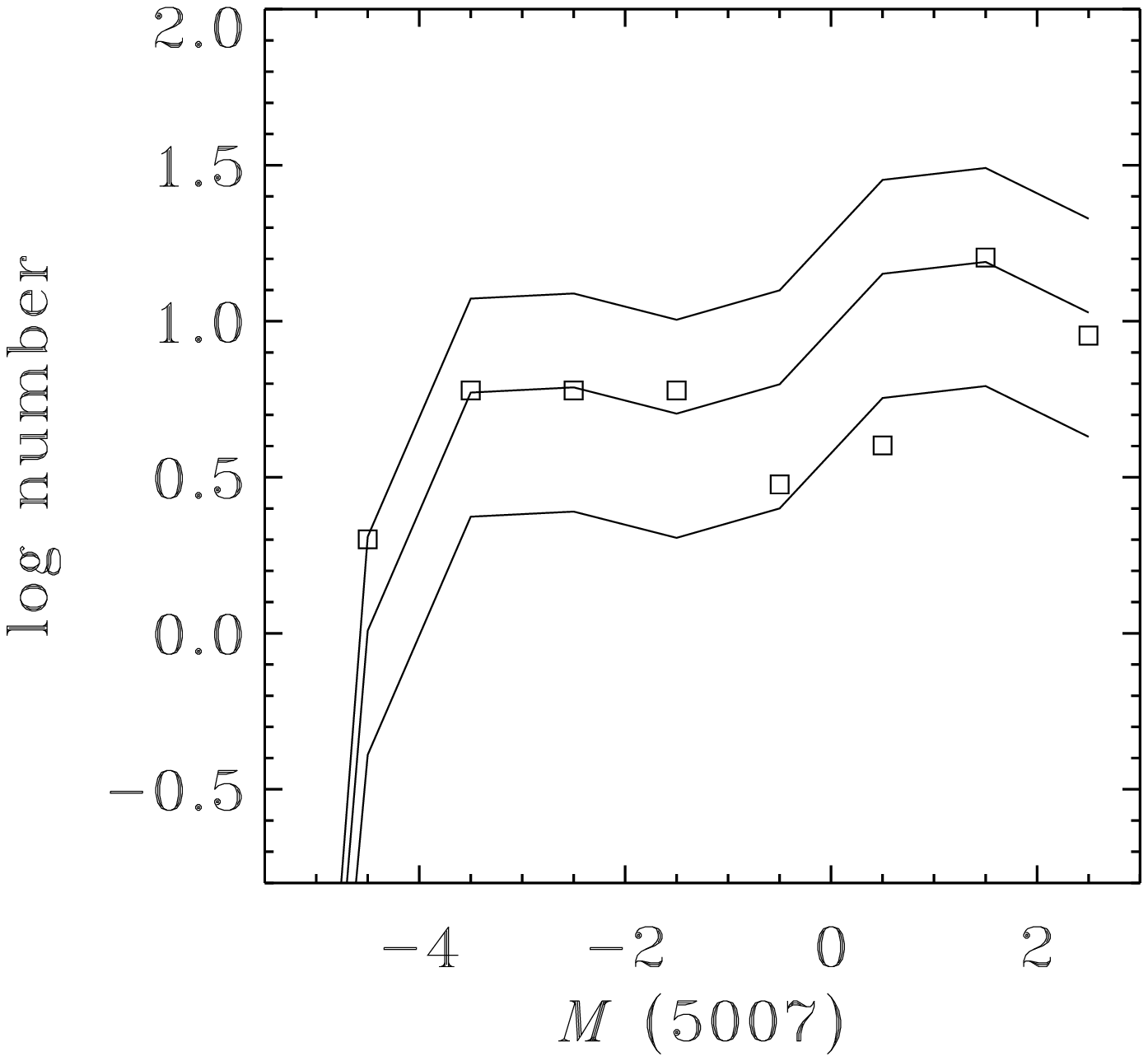}
\caption{
Fits to the 59 PNs found in the SMC by Jacoby and De Marco (2002) 
with the new PNLF simulations. The PNLF binning is broad (1 mag) 
due to the small number of PNs. The fainter bins are probably 
affected by incompleteness in the SMC surveys. The left figure, 
which provides a better fit, corresponds to the upper central star
mass distribution in Fig. 8, with the sharp low-mass cut at 0.585 
$M_{\odot}$. The fit was obtained adopting a distance modulus 
$m - M$ = 19.34. The three lines are PNLF simulations for
three different sample sizes: 70, 120, 190.  
The right figure corresponds to the 3rd central star mass 
distribution in Fig. 8. The fit was obtained adopting a distance 
modulus $m - M$ = 19.25, and the sample sizes are 30, 75 and 150.
In both figures we fit only the bright end of the PNLF, as 
discussed in the text.
}
\end{figure}


\begin{thebibliography}{}
\bibitem[]{583}
Acker, A., Ochsenbein, F., Stenholm, B., et al. 1992, The Strasbourg-ESO
                                              Catalogue of Galactic PNs
\bibitem[]{586}
Alves, D.R., Bond, H.E., \& Livio, M. 2000, AJ, 120, 2044
\bibitem[]{588}
Bl\"ocker, T. 1995, A\&A, 299, 755
\bibitem[]{590}
Ciardullo, R. 2003, in IAU Symposium 209, Planetary Nebulae: their 
evolution and role in the universe, ed. S. Kwok, M. Dopita \& R. 
                                Sutherland (San Francisco: ASP), 617
\bibitem[]{594}
Ciardullo, R. 2006, in IAU Symposium 234, Planetary Nebulae in our 
Galaxy and beyond, ed. M.J. Barlow \& R.H. M\'endez (Cambridge 
                                              University Press), 325
\bibitem[]{598}
Ciardullo, R., Durrell, P.R., Laychak, M.B., et al. 2004, ApJ, 614, 167
\bibitem[]{600}
Ciardullo, R., Feldmeier, J.J, Jacoby, G.H., et al. 2002, ApJ, 577, 31
\bibitem[]{602}
Ciardullo, R., Jacoby, G.H., Ford, H.C., \& Neill, J.D. 1989, ApJ, 339, 53
\bibitem[]{604}
Ciardullo, R., Sigurdsson, S., Feldmeier, J.J, \& Jacoby, G.H. 2005, ApJ,
                                                                629, 499
\bibitem[]{607}
Dopita, M.A., Jacoby, G.H., \& Vassiliadis, E. 1992, ApJ, 389, 27
\bibitem[]{609}
Ferrario, L., Wickramasinghe, D., Liebert, J., \& Williams, K.A. 2005,
                                                       MNRAS, 361, 1131
\bibitem[]{612}
Jacoby, G.H. 1989, ApJ, 339, 39
\bibitem[]{614}
Jacoby, G.H. \& Ciardullo, R. 1999, ApJ, 515, 169
\bibitem[]{616}
Jacoby, G.H. \& De Marco, O. 2002, AJ, 123, 269
\bibitem[]{618}
Jacoby, G.H., Walker, A.R., \& Ciardullo, R. 1990, ApJ, 365, 471
\bibitem[]{620}
Magrini, L., Corradi, R.L.M., Mampaso, A., \& Perinotto, M. 2000, A\&A, 
                                                                 355, 713
\bibitem[]{623}
Marigo, P., Girardi, L., Weiss, A., et al. 2004, A\&A, 423, 995
\bibitem[]{625}
McCarthy, J.K., M\'endez, R.H., Becker, S., et al. 1997, in IAU 
Symposium 180, Planetary Nebulae, ed. H.J. Habing \& H.J.G.L.M. 
                                       Lamers (Dordrecht: Kluwer), 122 
\bibitem[]{629}
Meatheringham, S.J., \& Dopita, M.A. 1991a, ApJS, 75, 407
\bibitem[]{631}
Meatheringham, S.J., \& Dopita, M.A. 1991b, ApJS, 76, 1085
\bibitem[]{633}
Meatheringham, S.J., Dopita, M.A., \& Morgan, D.H. 1988, ApJ, 329, 166
\bibitem[]{635}
M\'endez, R.H. 1999, in Post-Hipparcos Cosmic Candles,, ed. A. Heck \&
                                      F. Caputo (Dordrecht:Kluwer), 161
\bibitem[]{638}
M\'endez, R.H., Kudritzki, R.P., Ciardullo, R., \& Jacoby, G.H. 1993,
                                   A\&A, 275, 534
\bibitem[]{641}
M\'endez, R.H., Kudritzki, R.P., \& Herrero, A. 1992, A\&A, 260, 329
\bibitem[]{644}
M\'endez, R.H., \& Soffner, T. 1997, A\&A, 321, 898
\bibitem[]{646}
M\'endez, R.H., Thomas, D., Saglia, R.P., et al. 2005, ApJ, 627, 767
\bibitem[]{648}
Meng, X., Chen, X., \& Han, Z. 2007, astro-ph 0710.2397
\bibitem[]{650}
Perinotto, M., Kifonidis, K., Sch\"onberner, D., \& Marten, H. 1998,
                                                       A\&A, 332, 1044
\bibitem[]{653}
Sch\"onberner, D. 1989, in IAU Symposium 131, Planetary Nebulae, ed.
                             S. Torres-Peimbert (Dordrecht: Kluwer), 463
\bibitem[]{656}
Sch\"onberner, D., Jacob, R., Steffen, M., \& Sandin, C. 2007, A\&A, 473, 467
\bibitem[]{658}
van der Sluys, M.V., Verbunt, F., \& Pols, O.R. 2005, A\&A, 431, 647
\bibitem[]{660}
Vassiliadis, E., Dopita, M.A., Morgan, D.H., \& Bell, J.F. 1992, ApJS, 83, 87
\bibitem[]{662}
Weidemann, V. 2000, A\&A, 363, 647
\bibitem[]{664}
Wood, P.R., Meatheringham, S.J., Dopita, M.A., \& Morgan, D.H. 1987,
                                                         ApJ, 320, 178  
\end{thebibliography}
\end{document}